\documentclass[runningheads]{llncs}
\usepackage{graphicx}
\usepackage{cite}
\usepackage{hyperref}
\usepackage{amsmath}
\usepackage{amsfonts} 
\usepackage{svg}
\usepackage{amssymb}
\begin{document}
\title{Data-Enhanced Process Models in \\ Process Mining}
%
%
\author{Jonas Cremerius, Mathias Weske}
%
%
\institute{
Hasso Plattner Institute, University of Potsdam, Potsdam, Germany\\
\email{\{jonas.cremerius,mathias.weske\}@hpi.de} }
\maketitle              
\begin{abstract}
Understanding and improving business processes have become important success factors for organizations. Process mining has proven very successful with a variety of methods and techniques, including discovering process models based on event logs. Process mining has traditionally focussed on control flow and timing aspects. However, getting insights about a process is not only based on activities and their orderings, but also on the data generated and manipulated during process executions. Today, almost every process activity generates data; these data do not play the role in process mining that it deserves. This paper introduces a visualization technique for enhancing discovered process models with domain data, thereby allowing data-based exploration of processes. Data-enhanced process models aim at supporting domain experts to explore the process, where they can select attributes of interest and observe their influence on the process. The visualization technique is illustrated by the MIMIC-IV real-world data set on hospitalizations in the US.

\end{abstract}
\section{Introduction}

Business process models play a central role in exploring and analysing the execution of business processes in organizations. Using process mining methods and techniques, process models can be discovered from real-world process executions, using event logs~\cite{PMAction}. 

\begin{figure}
    \centering
    \includegraphics[width=12cm]{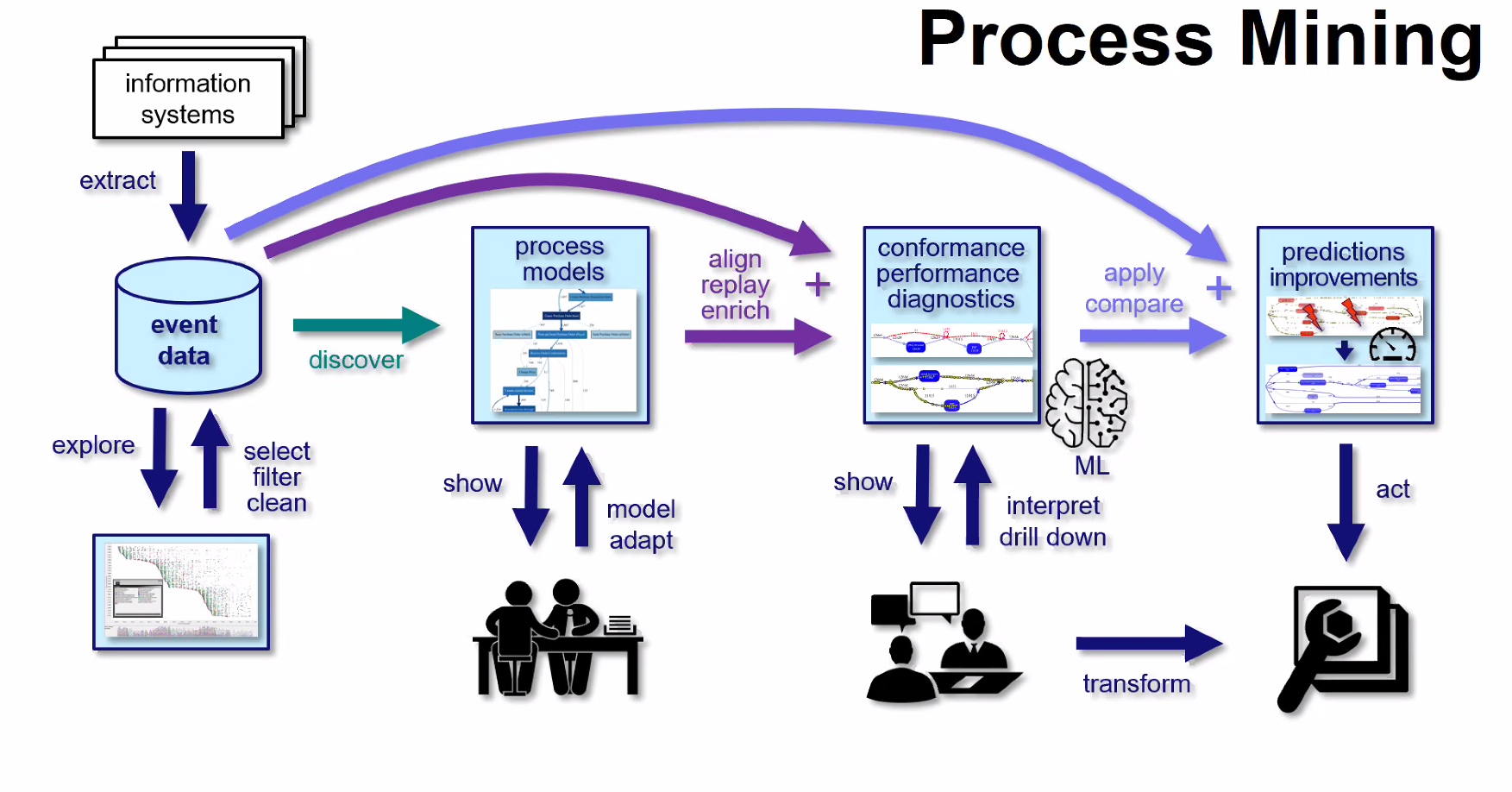}
    \caption{Main process mining steps \cite{wil_2021}}
    \label{fig:pm}
\end{figure}

Process mining supports a series of steps, which are illustrated in Fig.~\ref{fig:pm}. After an event log has been extracted from an information system, the discovered process model illustrates the order of activities represented in the respective event log. In comparison to other data analytic disciplines, process mining has the advantage of generating graphical representations, i.e., process models, which are helpful for communicating results to domain experts. 

With process models, domain experts can identify insights about the activities of a process, their order and frequency. However, an important element of any business process is not well represented in process models: domain data. Domain data refers to all data values generated throughout a process execution, which is represented as additional event/trace attributes in process mining besides the essential ones (case identifier, activity name, and timestamp). Domain data is an essential resource to understand how a process is actually conducted, and which effect on the process certain data values actually have. This contribution deals with the problem that domain data cannot be linked to the respective process activity in the process model where it was used or generated throughout the process.

This paper introduces data-enhanced process models that use domain data to enhance discovered process models. The visualization technique has been developed in the context of the healthcare domain, where in clinical patient treatment processes data (e.g., laboratory values) are of paramount importance. With domain data included in the process model, their development throughout the process can be analysed. While the approach is particularly useful in healthcare, it can also be applied to other domains where data plays an important role, for instance, logistics.

The remainder of this paper is organized as follows. Section~\ref{Motivating} presents a motivating example from the healthcare domain. Following that, Section~\ref{related} provides related work and Section~\ref{framework} introduces the foundation and analysis possibilities of data-enhanced process models. After that, Section~\ref{Eval} applies our idea to the motivating example and discusses the influence of it on the process model. Finally, Section~\ref{final} concludes this contribution.   

\section{Motivating Example}
\label{Motivating}

Processes play a key role in the healthcare sector in general and in clinical environments in particular, as hospitals are becoming increasingly aware of the need to understand and improve their processes \cite{PMHC}.

The definition and analysis of so-called clinical guidelines that recommend diagnostic and treatment related steps for specific diseases is a popular research field in process management. 
Process Mining helps to analyse the adherence and deviation of these guidelines in practice \cite{recom}. 

During the diagnosis and treatment, a large amount of data is generated by process activities, such as ``Analyse Laboratory Values'' and ``Analyse Chest X-Ray Image''. In the respective event log, for both events, different sets of attributes are generated, such as the number of abnormal laboratory values and the findings in the respective region of the chest \cite{HF_guide}.

\begin{figure}
    \centering
    \includegraphics[width=12cm]{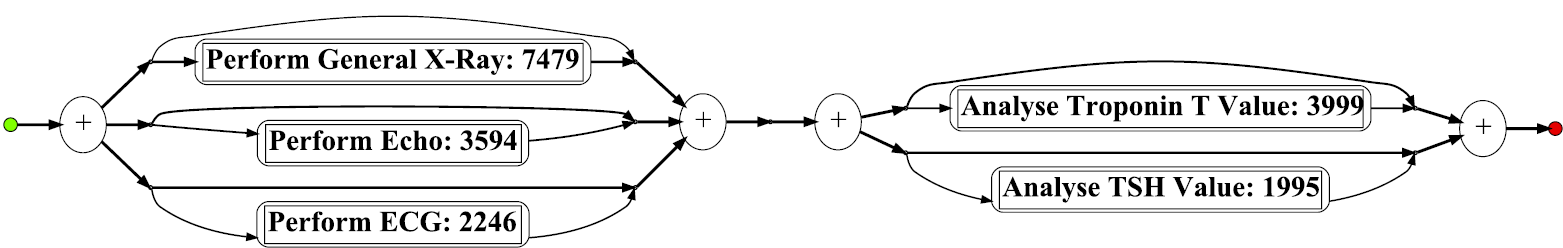}
    \caption{Process model for the diagnosis of heart failure, based on MIMIC-IV data set, showing the activities with their frequency. TSH: Thyroid Stimulating Hormone, ECG: Electrocardiogram}
    \label{fig:hf_1}
\end{figure}

An example diagnostic process of patients with acute heart failure is illustrated in Fig. \ref{fig:hf_1}, which is derived from an event log and was mined with the Inductive Visual Miner. The event log data is provided by the MIMIC-IV (v0.4) database, which contains information on hospital stays for patients admitted to a tertiary academic medical centre in Boston, MA, USA \cite{MIMIC}. Each activity describes the performed diagnostic method. At first, non-invasive methods are performed, i.e. ``Perform General X-Ray''. After that, different sets of laboratory values are analysed, such as ``Analyse TSH Value''. Then, the diagnostic process ends and the treatment starts. The process model does not include all diagnostic steps to keep the model simple and readable.

Based on that process model, the order and frequency of the diagnostic steps can be observed. It can be checked, for instance, if a recommended diagnostic step, as described in clinical guidelines, was actually performed. 

However, clinical guidelines describe not only the steps, but also statistical observations based on domain data values. For instance, in the vast majority of patients with acute heart failure, elevated concentrations of circulating cardiac troponins are detected. Or that the chest X-Ray in up to 20\% of patients with acute heart failure is nearly normal \cite{HF_guide}. 

Even though this information is essential to analyse the diagnostic process, it is not readily available in the discovered process models. Rather, it needs to be manually retrieved from event logs, if available. Hence, the process analyst needs to look into the event log and performs aggregations over the desired values manually.

As a result of this manual effort, the resulting values cannot be linked back to the process model and need to be presented separately, even though these values refer directly to the activities performed during the process. With the data presented separately in tables or in other forms, there is no link between the data behind the process and the process model. This contribution proposes a functionality to create this link in enabling data visualization inside the process model.

\section{Related Work}
\label{related}

Research in process mining mainly focusses on control-flow aspects, but there exist some approaches considering the data-oriented perspective. For instance, the multi-perspective process explorer makes use of data attributes attached to events showing the value distribution at certain states within the process model, which is provided as an additional tool besides the process model \cite{process-explorer}. Furthermore, data attributes are used to identify constraints at decision points, which are illustrated in the process model \cite{rules_1}. In the field of declarative process models, data attributes are used to provide a summarized view of key rules governing the process, where the rules are displayed at the edges of the declarative process model \cite{declarative_PM}.

Value stream mapping (VSM) is a technique for the diagnosis, implementation, and maintenance of a lean approach. VSM visually display the flow of materials and information, including data. Process activities in VSM can be enhanced by data tables, including information about the cycle time, reliability, number of employees, and more \cite{VSM_1, VSM_2}.

Inspired by VSM, this contribution extends the usage of data attributes by enabling monitoring of these inside the process model and bringing the event attribute values to the place where they have been used or modified, which is the process activity representing events with the same activity name.

\section{Data-Enhanced Process Models}
\label{framework}

This section introduces a visualization technique for enhancing discovered process models with domain data. The discussion starts with an event log containing the mandatory entries including case identifier, activity name, and timestamp. If an event contains additional information, it is included as an event attribute, which is defined by the XES standard \cite{XES_Standard}. 

\begin{figure}
    \centering
    \includegraphics[width=12cm]{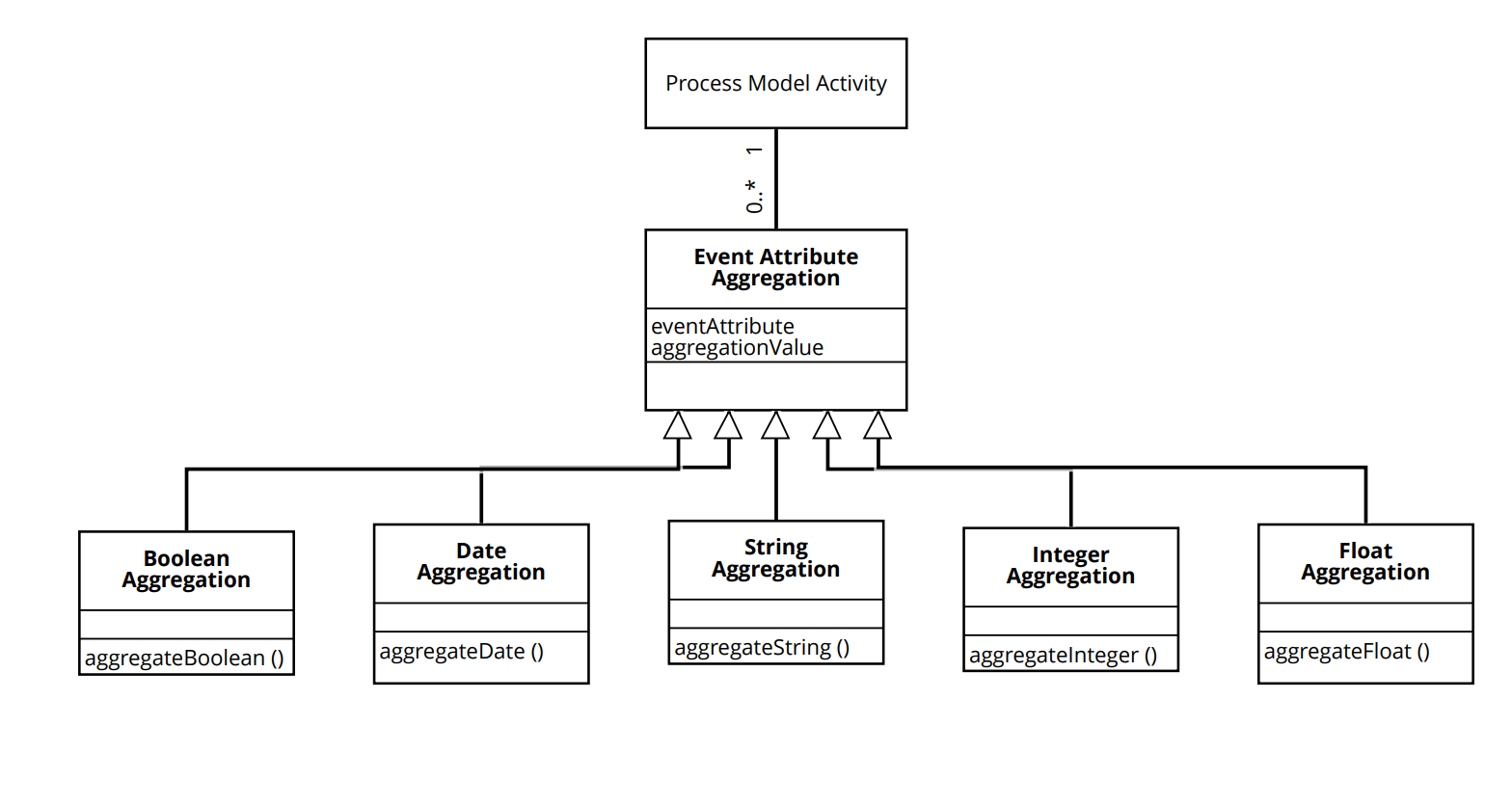}
    \caption{UML class diagram for adding event attribute aggregations to process model activities}
    \label{fig:class}
\end{figure}

Elementary event attributes have one of five data types: String, Date, Integer, Float, or Boolean. After the process model has been discovered, the activities in the process model can be graphically extended by their respective event attributes, as illustrated in Fig. \ref{fig:class}. Activities in the process model are different to events in the event log, as an activity in the process model represents all events with the same activity name in the event log. Thus, the process model represents an aggregation of all events with the same activity name, describing the general behaviour of them.

Therefore, data-enhanced process models use the concept of aggregation as well by aggregating event attribute values for presentation in the process model. Additionally, VSM use the concept of aggregation to display data values, e.g., reliability \cite{VSM_2}. This contribution introduces event attribute aggregations, which consist of the respective event attribute and the aggregated value, such as min, max, and mean \cite{aggregation}. 

\subsection{Foundation}

Event logs are the main input for this contribution. It consists of traces, which are sequences of events, where an event can have any number of event attributes. The following definition is based on~\cite{correlation}.

\subsubsection{Definition 1 (Event log, Trace, and Event)}

Let $V$ be the universe of all possible values and $E_{A}$ be the universe of event attributes. An event $e$ is a mapping of event attributes to values, such as $e \in E_{A} \to V$. The universe of events is defined as $E_{U} = E_{A} \to V$. A trace $t \in {E_{U}}^\star$ is a sequence of events, and $T = {E_{U}}^\star$ represents the respective universe of traces. An event log $L$ is a multi-set of traces, so $L \in M(T)$, where, given any set X, $M(X)$ is the set of all possible multi-sets over $X$.
\newline

\noindent
For each event, the event attributes case identifier, activity, and timestamp are required, as they are mandatory in the XES standard \cite{XES_Standard}. For an event $e$ these attributes are represented by $e(case\ identifier)$, $e(activity)$, and $e(timestamp)\neq \bot$, respectively, where $\bot$ indicates a null value.

\subsubsection{Definition 2 (Process Model)} Given an event log $L \in M(T)$ and a process discovery algorithm $\alpha$ taking $L$ as its input, then $\alpha(L)$ generates a process model $P = (N, E)$. $N$ is the set of process activities and represents the events in the event log, such that given an event $e$ in the log $L$, $e(activity) \in N$. $E$ is a set of directed edges, such that $E \subseteq N \times N$. Gateways or other model-specific elements are not considered in this definition, as they are not affected by data-enhanced process models.

We now define process variants, based on \cite{variants}.

\subsubsection{Definition 3 (Process Variants)}
 An event log $L \in M(T)$ can be partitioned into a finite set of groups called process variants $v_{1}, v_{2},...,v_{n}$,
 and there exists an attribute $d$, for which all trace variants have the same value.
 
\noindent
This definition ensures that the process variants share an event attribute value, so that each trace $t \in L$ belongs to one and only one process variant. For example, given two events $e_{k}$, $e_{l}$ and a process variant $v_{i}$ which is filtered based on $age$, so that $d = age$, $e_{k}$ and $e_{l}$ belong to $v_{i}$, if and only if $e_{k}(age) = e_{l}(age)$. As a process variant is a subset of an event log, activities could be completely omitted when the process variant is analysed in the general process model of the whole event log.

Based on the observation that each activity in a process model represents many events, each of which have dedicated attribute values, we define aggregation operations. These operations aggregate attribute values of all events that are represented by an activity in a discovered process model.

\subsubsection{Definition 4 (Event Attribute Aggregation)} An event attribute aggregation is a tuple $e_{AA} = (a, e_{At}, E_{AV}, \lambda_{A})$ consisting of

\begin{itemize}
  \item an activity $a \in N$ for which the event attributes are extracted
  \item an event attribute $e_{At} \in E_A$, which is aggregated
  \item a multi-set with all event attribute values to be aggregated $E_{AV} \subseteq V$
  \item an aggregation function $\lambda_{A} = E_{AV} \to \mathbb{Q}$
\end{itemize}
Given an event log $L$, the desired event attribute values can be extracted by iterating through each trace $t \in L$, including the contained events:
\begin{equation}
    E_{AV} := \{e(e_{At}) \in V \mid e(activity) = a, e \in t, t \in L \}
\end{equation}

\noindent
With $E_{AV}$ representing all event attribute values for the selected activity, the aggregation function can be applied. Therefore, $\lambda_{A} = E_{AV} \to \mathbb{Q} $ takes the values in $E_{AV}$ and aggregates them to one number. The only requirement for a trace to be included in the aggregation calculation is the occurrence of the event with the respective activity name. Therefore, if a trace is not fully compliant with the model, the trace will still be considered in the aggregation. If an event with the same activity name is executed multiple times in a trace, all respective event attribute values will be used in the aggregation function. 
Suggested aggregation functions are described in Section \ref{aggregations}, such as mean, median, and frequency.

The event attribute aggregations can be attached to the process model, creating a data-enhanced process model.

\subsubsection{Definition 5 (Data Enhanced Process Model)}
Let $E_{AA}$ be the universe of event attribute aggregations. An event attribute aggregation $e_{AA}$ is a quadruple, as defined in Definition 4, such that $e_{AA} \in E_{AA}$. Given a process model $P = (N, E)$, a data-enhanced process model $DEP$ enhances each activity $N$ by a set of event attribute aggregations, resulting in $N_{DEP} \subseteq (N \times E_{AA})$. The data-enhanced model is then a tuple of data-enhanced activities $N_{DEP}$ and edges $E$, such that $DEP = (N_{DEP}, E)$.
\newline
\newline
\noindent
By enhancing the process model activities by event attribute aggregations, the link between the process model and event log is created. For each process model activity $a \in N$, multiple event attribute aggregations can be added.

\subsection{Event Attribute Selection and Aggregation}

The event attribute selection is supposed to be flexible in the sense of being able to add and remove event attribute aggregations while analysing the process together with a domain expert. As the domain expert knows best which event attributes could have an influence on an activity, the flexibility of adding and removing event attribute aggregations is key in enabling an interactive, exploratory process mining experience.

The possible aggregation methods are provided for each data type separately. This contribution differentiates between categorical, continuous, and discrete variables. Categorical variables contain a finite number of distinct groups, which is the case for the data types String and Boolean. However, the data types Date, Integer, and Float could be categorical as well, if the number of distinct groups is comprehensible. For example, a Boolean might be represented as an Integer, where True and False are stored as 0 and 1 respectively. Integer is a data type with a countable number of values between any two numbers and is therefore discrete. Date and float have an infinite number of values between any two values, which makes them continuous \cite{math}.

\label{aggregations}
The following aggregation functions ($\lambda_{A}$) are suggested for continuous and discrete variables \cite{aggregation}.
The minimum and maximum can be used to observe the value range of event attributes. The arithmetic mean provides an averaging aggregation function considering each value in the data-set equally. It is derived from the sum of all values divided by the number of values.
The median in a set of values is defined as the value in the middle, once the values are ordered. In comparison to the arithmetic mean, the median ignores outliers, but considers a subset of the values only.

As the aggregation functions introduced so far cannot be applied for categorical variables, we propose additional aggregation functions for those \cite{categorical}.
The frequency can be used to count the occurrences of an event attribute value. It gives an understanding of the absolute occurrence of a value in process instances.
The percentage or proportion of a value in a set of values can be monitored in a process model. These are often referred to as the relative frequency.

With these aggregation functions at hand, determining the most appropriate aggregation method is addressed next. It is recommended to involve the domain expert in choosing an aggregation method by asking for the desired output. The active involvement of domain experts strengthens the interactive process mining experience, as proposed in Fig. \ref{fig:pm} \cite{aggregation}. Besides that, the process analyst should ensure that the aggregation method is as simple and as understandable as possible to reach the desired goal.

The listing of aggregation functions in this contribution is not excessive, as the possible number of aggregation functions is infinite \cite{aggregation}. It just covers the mostly used and appropriate ones for inclusion in the process model.

\subsection{Process Analysis Possibilities}
Data-enhanced process models enable to analyse processes together with a domain expert in the following ways.

\begin{itemize}
\label{use case 1}
    \item \textit{Monitoring event attribute values}: At first, the values of the event attribute aggregations can be monitored in the process model. For instance, it can be checked, if attributes exceed given values at a given process step. These are typically defined by process guidelines; for example, the adherence to admission guidelines for the intensive care unit could be monitored \cite{icu_guide}. Furthermore, the change of event attribute values during process execution could be observed, if activities share the same event attributes. Examples include the progress of a parcel delivery or a patient's blood pressure throughout a hospital admission.
    \item \textit{Comparison of process activities}: In process models, different activities can help to reach the desired goal. To which extent these activities actually help to reach the goal can be measured by extracting the respective event attributes, i.e., the percentage of successfully completing the activity or the resource efficiency (time, workforce, material, etc.). Showing these measures in the process model makes the activities comparable not only based on their control flow, but also on their event attribute values. For example, an activity might be performed frequently, but does not contribute to the business goal and consumes more resources than other activities having a higher impact on the business goal. Considering a parcel delivery process, different means of transportation could be compared by their resource efficiency.
    \item \textit{Exploration and comparison of process variants}: Integrating event attributes in the process model can also help to explore process variants with the aim of generating insights or confirming expected process behaviour. Process variant analysis tries to identify differences existing in a set of process executions enacted by the same process model \cite{variants}. Given an event log $L$ and a resulting data enhanced process model $DEP$, the event attribute values of a given process variant $v_{i} \subseteq L$ can be monitored in $DEP$. As the event attribute aggregations are already attached to $DEP$, the changes in the event attributes can be observed without having to recalculate the statistics manually. Hence, it allows to compare the values of the event attributes immediately after filtering the events accordingly. Furthermore, it is still possible to add event attribute aggregations of interest for the specific process variant $v_{i}$. As a process variant $v_{i}$ is a subset of the event log $L$, it might not cover all process activities, so that some activities including their respective event attribute aggregations could be omitted. An example process variant comparison could be performed for the treatment process of patients with different diseases. Then, not only the activities and control flow, but also the development of the patient state and further attributes can be compared directly in the process model.
\end{itemize}

\section{Experiment and Discussion}
\label{Eval}

We start the experiment by considering the main steps in process mining illustrated in Fig. \ref{fig:pm}, where the touchpoint of this contribution can be observed. First, an event log $L$ is extracted from an information system. Then, a process model $P$ is mined from $L$. After that, the set of activities $N$ are enriched by event attribute aggregations $E_{AA}$, resulting in a data enhanced process model $DEP = (N_{DEP}, V)$. It is important to note, that event attribute aggregations $e_{AA}$ can be added on demand, creating new pairs $(N_{DEP} \times E_{AA})$, while analysing the process model $DEP$. Therefore, the core of data-enhanced process models is at the enrichment step, as illustrated by the purple arrow in Fig. \ref{fig:pm}. From there, the analysis of event attributes inside the process model can be performed.

The proposed visualization technique is implemented as an extension of a plug-in in ProM. The prototype implementation extends the Inductive Visual Miner\footnote{\url{https://svn.win.tue.nl/repos/prom/Packages/InductiveVisualMiner}}. It is tailored to the event log for heart failure patients, which can be created by executing a script for event log generation, which has been developed by the authors. The prototype implementation and the script can be downloaded\footnote{\url{https://github.com/jcremerius/Data-Enhanced-Process-Models}}. Thus, the main input is an event log $L$ of which a process model is mined by the Inductive Visual Miner. The event log was derived from the MIMIC-IV database \cite{MIMIC}, which, as introduced in Section \ref{Motivating}, records real-world hospital data. MIMIC-IV incorporates the whole patient journey through the hospital, including procedures performed, medications given, lab values taken, image analysis conducted, and more. Thus, it provides additional data for activities performed in a process.

\begin{figure}
    \centering
    \includegraphics[width=12cm]{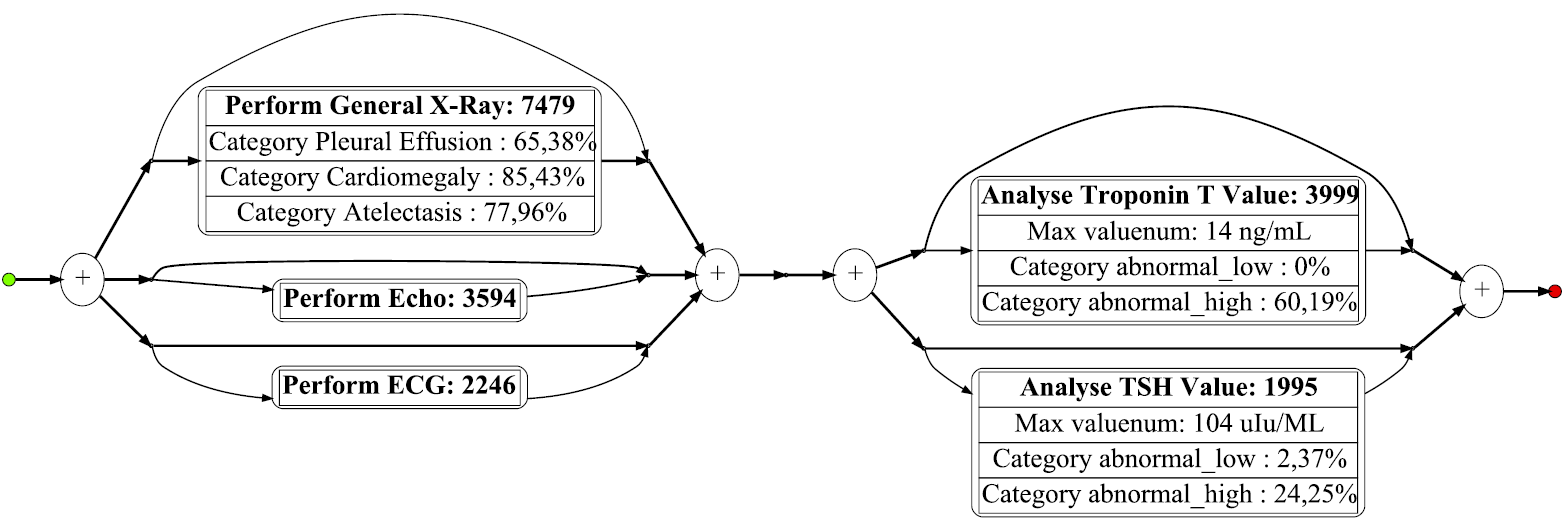}
    \caption{Data enhanced process model for the diagnosis of heart failure. TSH: Thyroid Stimulating Hormone, ECG: Electrocardiogram}
    \label{fig:hf_guide_simple}
\end{figure}

To illustrate the approach, a sample process model is displayed in Fig. \ref{fig:hf_guide_simple}, which shows the process model illustrated in Fig. \ref{fig:hf_1}, with event attributes attached to some activities. In particular, the activity ``Perform General X-Ray'' is extended by the percentage of malignant findings, including Pleural Effusion, Cardiomegaly, and Atelectasis. Therefore, this activity includes three event attribute aggregations, one for each malignant finding, resulting in three pairs in the relation $(N_{DEP} \times E_{AA})$. The activities ``Analyse X Value'' are extended by the percentage of abnormalities observed and the maximum laboratory value.

Data-enhanced process models allow us to compare treatment processes with clinical guidelines~\cite{HF_guide}. For instance, a clinical guideline states that elevated concentrations of circulating cardiac troponins are detected in the vast majority of patients with acute heart failure and that the chest X-Ray in up to 20\% of the patients is nearly normal. This property can now be evaluated by exploring the process model. In particular, looking at the activity ``Analyse Troponin T Value'', the elevated concentration can be observed by checking abnormally high cardiac troponin levels, which occur in 60\% of all patients. The frequency of findings in an X-Ray can also be observed in the activity ``Perform General X-Ray''. 
This type of analysis refers to the first analysis possibility introduced above (monitoring event attribute values). Additionally, novel characteristics can be derived, which are not yet stated in clinical guidelines, such as that the Thyroid Stimulating Hormone is more often abnormally high than low \cite{HF_guide}.

Regarding the comparison of process activities, the activities that analyse lab values can be compared by their frequency of finding an abnormal lab value. That could help to assess the efficiency of taking a specific lab value.

The third analysis possibility (exploration and comparison of process variants) could be applied by investigating the subset of patients having an abnormal Troponin T value. Then not only the changes in the control flow, but also the changes in the event attribute aggregations can be observed.

\clearpage

This contribution recommends bringing data directly into process models in the form of event attribute aggregations. We do not propose novel process analysis techniques, but suggest a new perspective on process models in process mining. With these data-enhanced process models, we can explain the behaviour of processes in different domains to confirm expected behaviour and to find novel, data-based insights. 

This opens the question, which role do domain experts play in process analysis. As illustrated in Fig. \ref{fig:pm}, the analysis should be conducted in an interactive, exploratory manner, as the domain expert knows best, what is of interest to her. However, the domain expert is bound to the information which is presented to her. Therefore, the goal of this approach is to empower the domain expert to participate in the analysis by choosing the attributes of interest in the process model. This helps to ensure that the information presented is tailored to the needs of the domain expert.

Representing activities solely by their activity name in a process model might not be enough for future process analysis. Big data systems deliver a detailed view of the real world, providing additional information for process activities.
We already see that domain data is analysed separately, making the process model more or less obsolete. Therefore, the relevance of process models needs to be kept in sight, which has the advantage of being able to show the control flow. Including additional data strengthens that advantage and allows to observe the interplay of process activities and domain data.

\section{Conclusion and Future Work}
\label{final}

This paper discusses the possibilities of including event attribute values directly into discovered process models. With the increasing data availability, a more comprehensive view of the process is possible, where the change of event attribute values during process execution can be observed. As process models can be used as a means of communication between process analysts and domain experts, incorporating event data in the process model has the potential to improve that communication by illustrating the control flow and event attributes in one place.

Based on this contribution, we could design an event attribute recommendation system, which identifies interesting event attributes based on their distribution or behaviour throughout the process. Furthermore, similar activities in regard to their event attributes could be automatically identified for comparison. Lastly, the role of domain experts in analysing business processes could be empirically investigated.  
\clearpage

%
%
%
 \bibliographystyle{splncs04}
 \bibliography{references}

\end{document}